\documentclass{PoS}

\title{Strong CP problem, axions, and cosmological implications  of CP violation}

\ShortTitle{CP and Cosmology}

\author{\speaker{Jihn E. Kim}\thanks{This work is supported in part by the National Research Foundation (NRF) grant funded by the Korean Government (MEST) (NRF-2015R1D1A1A01058449) and  the IBS (IBS-R017-D1-2016-a00).}\\
        Center for Axion and Precision Physics Research (CAPP, IBS),
  291 Daehakro, Yuseong-Gu, Daejeon 34141, Republic of Korea, and \\
     Department of Physics, Kyung Hee University, 26 Gyungheedaero, Dongdaemun-Gu, Seoul 02447, Republic of Korea  \\   E-mail: \email{jihnekim@gmail.com}}

\abstract{In this talk, I  present a pedagogical and historical review related to the CP symmetry.   
 It includes the weak CP violation, the strong CP problem, ``invisible'' axions and cosmology, and Type-II leptogenesis.  
}

\FullConference{Corfu Summer Institute 2016\\	31 Aug.-12 Sep. 2016\\ Corfu, Greece}

\newcommand\etal{{\it et al.}}
 \newcommand\ie{{\it i.e.}~}

\newcommand\vew{v_{\rm ew}}
\newcommand\Mp{M_{\rm P}}

\newcommand\UPQ{U(1)$_{\rm PQ}$}
\newcommand\UG{U(1)$_{\rm global}$}
\newcommand\Uanom{U(1)$_{\rm anom}$}
\newcommand\NDW{N_{\rm DW}}
   
 \newcommand{\dell}{\delta_{\rm PMNS}}
\newcommand{\delq}{\delta_{\rm CKM}}

\newcommand\gev{\,{\rm GeV}}

\newcommand{\Z}{{\bf Z}}

\begin{document}
  
 \maketitle

\section*{}\label{sect:intro}
\vskip -1.5cm
 
\section{Introduction}\label{sect:intro}
\begin{figure}[!h]
  \begin{center}
 \includegraphics[width=0.5\textwidth]{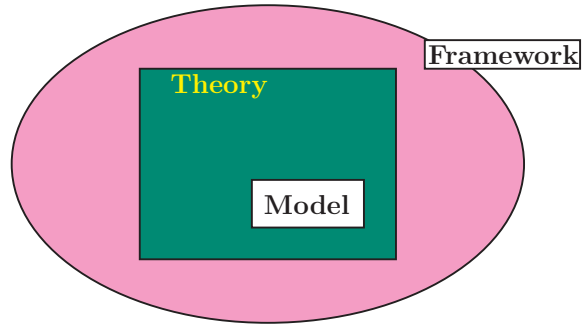}
  \end{center}
 \caption{An inclusive relation of frameworks due to Gross \cite{Gross16}.
  }
\label{fig:Gross}
\end{figure}

From a fundamental point of view,  presumably the most basic parameters are given at a mass-defining scale which is considered to be the Planck mass $\Mp\simeq 2.43\times 10^{18\,}\gev$. Any other scale involves a small coupling which is that scale divided by $\Mp$. Below $\Mp$, one tries to understand elementary particles in this sense. 
Figure \ref{fig:Gross} is a theorist's design to plan his idea, starting with a grand framework, presumably inclusive of as much natural phenomena as possible. Within this framework, he builds a theory. The   theory contains models.  These models must be working examples, explaining the observed phenomena. Without a model example, some will say that it is a religion even though the design is fantastic. Our job is to find  working models in this FRAMEWORK/THEORY/MODEL  inclusion cartoon. In this sense, ``symmetry'' as a framework has worked for a century.

Nowadays, flavor symmetries are studied mainly by some discrete symmetries, due to the observed large mixing angles in the leptonic sector. But, flavor symmetry may be a gauge symmetry in which case a true unification is GUTs with the flavor symmetry included there. The first attempt along this line was due to Georgi in SU(11) \cite{Georgi79} on the unification of GUT families (UGUTF). The next try with spinor representation of SO($4n+2$) groups was in SO(14) \cite{KimPRL80}. A more attempt along this line is  from string compactification  based on ${\bf Z}_{12-I}$ orbifold compactification \cite{KimJHEP15}. But, mostly one tries to obtain an electroweak-scale massive particle by a discrete
symmetry and a light scalar by the Goldstone theorem.
Among Goldstone bosons,
``invisible'' axion is the most interesting one.

\section{CP's}

To discuss violation of a symmetry, first one has to define the symmetry. Parity  P  is the most well-known example for the definition and violation of a symmetry. Even though kinetic mixings of U(1) gauge bosons have been considered for some time, the definition of a symmetry is usually done such that the kinetic energy terms preserve the symmetry. 
If there exists a possibility of a Lagrangian that satisfies $(\rm CP){\cal L}(CP)^{-1}={\cal L}$, then the CP symmetry is preserved. Here,
the first thing to do is to define fields with CP quantum numbers. Next, find out terms breaking CP and search for its physical implications.

CP violation is an interference phenomenon. At this workshop, Domencico \cite{OkunCited} cited Okun's statement, ``Neutral K mesons are a unique physical system which appears to be created by nature to demonstrate, in the most impressive manner, a number of most spectacular phenomenon''. One may replace (spectacular)$\to$(interference).
 CP violation in the SM is an interference phenomenon, encompassing all three families. This will become clearer below when we express the Jarlskog determinant $J$.
 
After the discovery of weak CP violation in 1964, needs for 
theories of weak CP violation  became very important. Some of these weak CP violation were
\begin{eqnarray}
&1.& \textrm{by light colored scalars \cite{KM73},}\nonumber \\
&2.& \textrm{by right-handed current(s) \cite{Mohapatra72,KM73},}\nonumber  \\ 
&3.& \textrm{by three left-handed families \cite{KM73}} \\ 
&4.& \textrm{by propagators of light color-singlet scalars \cite{Weinberg76},}\nonumber  \\ 
&5.& \textrm{by an extra-U(1) gauge interaction \cite{Wolfenstein64},}  \nonumber 
\end{eqnarray}
where some examples interpretable in modern gauge theories are cited. Item 3 is known as the Cabibbo\cite{Cabibbo63}-Kobayashi-Maskwa(CKM) model.
 
In the standard model(SM), the kinetic energy terms of quarks, leptons, and Higgs doublets are CP conserving. The CP violation in the SM arises in the interaction terms, typically through the Yukawa couplings. If the VEVs of Higgs doublets vanish, then there is no CP violation because all fermions are massless. Below the VEV scale of the Higgs doublets, all the SM fields obtain masses, and one can locate the CP phase in the left-handed currents, coupling to $W^\pm_\mu$.  The charged current couplings are defined in this setup for  the CKM (for quarks) and Pontecorvo-Maki-Nakagawa-Sakata(PMNS) (for  leptons) \cite{PMNS} matrices.
The CKM  matrix is unitary, which is the only condition for the CKM  matrix.  The physical significance of the weak CP violation is given by the  Jarlskog determinant $J$ which is obtained from the imaginary part of a product of two elements of $V$ and two elements of $V^*$ of the CKM matrix, e.g. of the type $J=|{\rm Im}(V_{12}V_{23}V_{13}^*V_{22}^*)|$ \cite{Jarlskog85}.

Let us choose the CKM and PMNS matrices such that the 1st row real.  Then, the CKM matrix can be chosen as \cite{KimSeo11},
\begin{eqnarray}
V_{\rm KS}= \left(\begin{array}{ccc} c_1,&s_1c_3,&s_1s_3 \\ [0.2em]
 -c_2s_1,&e^{-i\delq}s_2s_3 +c_1c_2c_3,&-e^{-i\delq} s_2c_3+c_1c_2s_3\\[0.2em]
-e^{i\delq} s_1s_2,&-c_2s_3 +c_1s_2c_3 e^{i\delq},& c_2c_3 +c_1s_2s_3 e^{i\delq}
\end{array}\right),~ {\rm~ with~Det\,}V_{\rm KS}=1 \label{eq:KSexact}
\end{eqnarray}
where $c_i=\cos\theta_i$ and $s_i=\sin\theta_i$ for $i=1,2,3$. In this form, the invariant quantity for the CP violation, the Jarlskog determinant is directly seen from $V_{\rm KS}$ itself \cite{KimNam15},\footnote{For the PMNS matrix, we use different parameters by replacing $\theta_i\to\Theta_i$ and $\delq\to \dell$.}
\begin{eqnarray}
J=|{\rm Im}\,V_{13} V_{22} V_{31} |\simeq O(\lambda^6),
\end{eqnarray}
where  $\lambda\simeq 0.22$ is the Cabibbo parameter $\lambda=\sin\theta_C$ \cite{Cabibbo63}. Note that all three families participate in the evaluartion of $J$, fulfilling the claim that CP violation is an `` interference phenomenon''. For the CP violation to be nonzero, in addition,
all u-type quark masses
must be different and all d-type
quark masses must be different, because one can rotate the phases of identical-mass quarks such that $\delq$ changes,  which implies that  $\delq$ is unobservable. 
  
\begin{figure}[t!]
 {\includegraphics[width=0.5
 \textwidth]{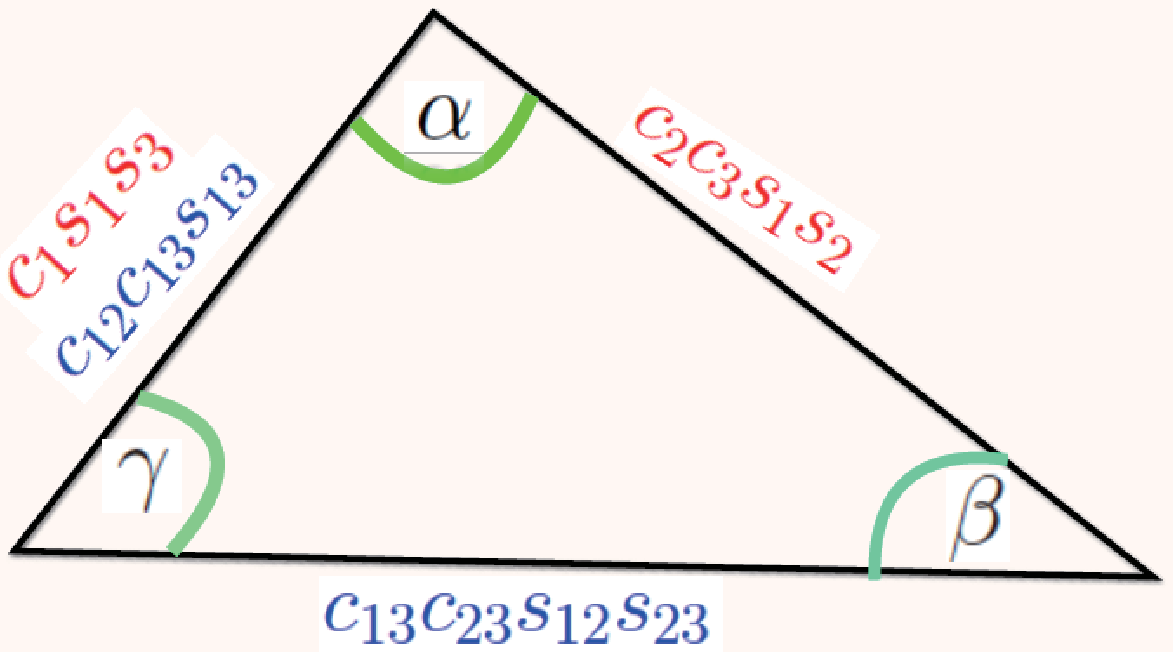}} \hskip 0.5cm  {\includegraphics[width=0.45
 \textwidth]{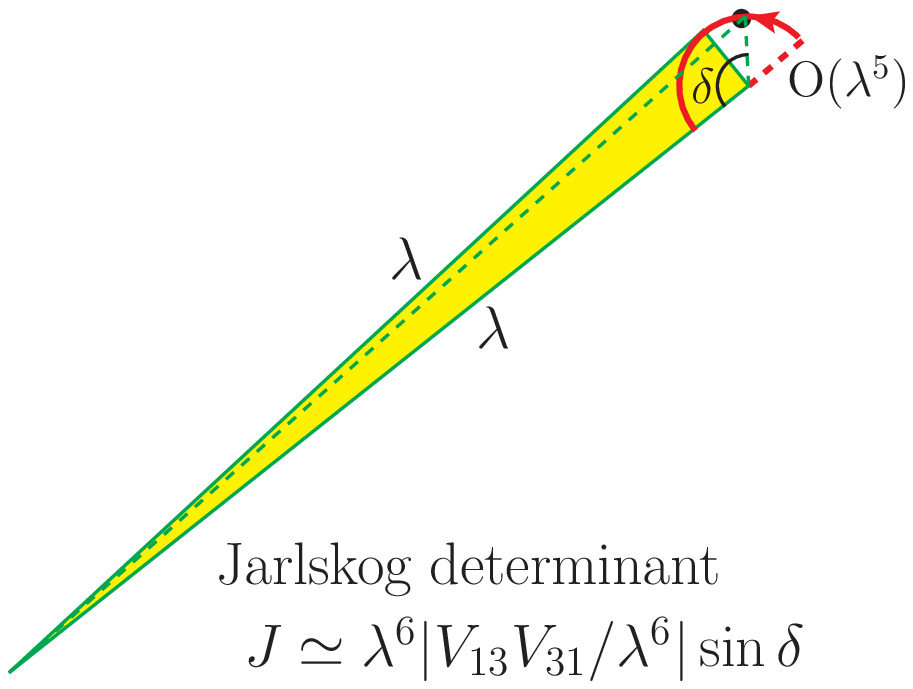}} \\
~ \hskip 3cm (a) \hskip 8.5cm (b)
\caption{The Jarlskog triangles. (a)   B-meson decay including a K-meson, and (b) B-meson decay to  $\pi$ mesons.}\label{Fig:Jarls}
 \end{figure}
  
There are many different parametrization schemes.  Different parametrizations give different CP phases $\delq$ \cite{KimNam15}. This can be understtod by the Jarlskog triangle for B-meson decay, shown in Fig. \ref{Fig:Jarls}\,(a).
For the parametrization of Eq. (\ref{eq:KSexact}), $J$ is given with the red parameters
\begin{equation}
J=c_1 c_2c_3 s_1^2 s_2 s_3 \sin\alpha
\end{equation}
while the PDG parametrization with the blue parameters fives $J=c_{12}c_{31}^2c_{23}s_{12} s_{23}s_{13}\sin\gamma$ \cite{PDG15}. In any parametrization, the area of Fig. \ref{Fig:Jarls}\,(a) gives the same value.

We can use the KS--parametrization, Eq. (\ref{eq:KSexact}), to show the maximality of the weak CP violation for the measured values of real angles $\theta_i$.  Let us  use the fact that any Jarlskog triangle has the same area. So, consider  Fig. \ref{Fig:Jarls}\,(b).
 With the $\lambda=\sin\theta_C$ expansion, the area of the Jarlskog triangle is of order $\lambda^6$, which is the product of two sides enclosing the angle $\delta$. Rotate the side of O($\lambda^5$) in the red arrow direction, making a triangle implied by dashed lines. This area of the triangle is maximal when the rotating angle $\delta\simeq 90^{\rm o}$, which is the triangle enclosing the yellow in  Fig. \ref{Fig:Jarls}\,(b). Note that $\alpha$ of   Fig. \ref{Fig:Jarls}\,(a) is determined by the fitting groups close to $\alpha= \frac\pi{2}$,
 \begin{eqnarray}
 \alpha=\left(85.4^{+3.9}_{-3.8}
 \right)^{\rm o}~[\rm PDG,\cite{PDG15}],~\left(88.6^{+3.3}_{-3.3}
 \right)^{\rm o}~[\rm U_{fit}, \cite{HarnewH16}],~\left(90.6^{+3.9}_{-1.1}
 \right)^{\rm o}~[\rm CKM_{fit},\cite{Pich16}].
 \end{eqnarray}
 With the KS parametrization,  $\delta\simeq \frac\pi{2}$ is in the allowed region. So, we proved that the weak CP violation is maximal with the pre-fixed real angles in  the KS--parametrization. Since physical statements are parametrization independent, this maximality must be the case in the PDG--parametrization also.
  
In the leptonic sector also, there is a preliminary hint that $\dell\ne 0$, and close to $-\frac{\pi}{2}$ even though the error bar is large \cite{T2KCP}. The quark mixing angles are  $\theta_i$ and $\delq$, and lepton mixing angles are  $\Theta_i$ and $\dell$.  
Even if  $\theta_i$ and $\Theta_i$ cannot be related, we can relate $\delq$ and $\dell$ if there is only one CP phase in the whole theory. Indeed, this has been shown in Ref. \cite{KimPLB11} where the weak CP violation is spontaneous and one unremovable phase is located at the weak interaction singlet {\it \`a la} the Froggatt-Nielsen(FN) mechanism \cite{FN79}.\footnote{After the talk, another method for the flavor solution has been proposed \cite{UanomFl}.} 
 In the supersymmetric model, it was shown that one phase in the ultra-violet completion gives \cite{KimNam15}:
$\dell=\pm\delq.$
Then, the Jarlskog triangles of the quark and lepton sectors will have one common phase.
   
In addition to the CKM and PMNS phases, there are Majorana and leptogenesis phases also. 
If there is only one phase in the ultra-violet completed theory, all of these must be expressed in terms of one phase.  So, the Majorana phase determined at the intermediate scale and the leptogenesis phase can be also expressed in terms of this one phase, as shown in Ref. \cite{CoviKim16}.

\section{The strong CP problem}

 Because of instanton solutions of QCD, there exists an effective
interaction term containing the gluon anomaly: $\bar{\theta}\{G\tilde{G}\}$. It is the flavor singlet and the source solving the U(1) problem of QCD by  't Hooft \cite{Hooft86}.
This gluon anomaly term is physical, but leads to
\begin{itemize}
\item
The strong CP problem,  ``Why is the nEDM so small?'' This leads to the three classes of natural solutions.
\item
The remaining `natural solution' is ``invisible'' axion \cite{KimRMP10}.
\end{itemize}
The three classes of solutions are (1) calculable models, (2) massless up quark, and (3) ``invisible'' axion. Calculable models is not   separable from the discussion of the weak CP violation.  
\begin{figure}[t!]
\centerline{\includegraphics[width=0.7\textwidth]{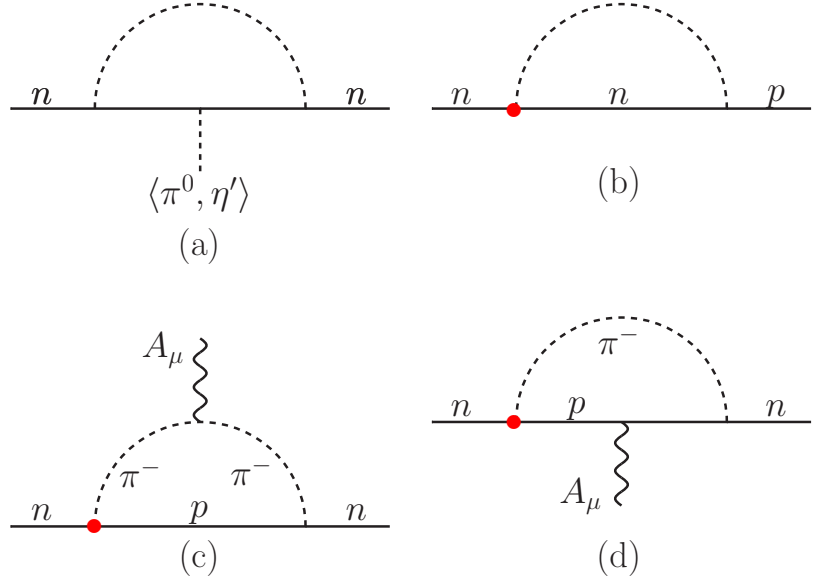}}
\caption{(a) CP violation by the insertion of $\langle \pi^0,\eta'\rangle$. The bullets in (b,c,d) are (a) of the CP violating interaction.  (c) and (d) lead  to an nEDM.}\label{Fig:nEDM}
\end{figure}
If the CP violating coupling $\overline{g_{\pi NN}}$ is present, the neutron electric dipole moment(nEDM) is calculated as  (with the CP conserving ${g_{\pi NN}}$ term),
\begin{equation}
\frac{d_n}{e}=\frac{{g_{\pi NN}}\overline{g_{\pi NN}}}{4\pi^2 m_N} \ln\left(\frac{m_N}{m_\pi}\right).\label{eq:nEDM}
\end{equation}
Figure \ref{Fig:nEDM}\,(a) shows $\overline{g_{\pi NN}}$, and Figure \ref{Fig:nEDM}\,(c,d) give  Eq. (\ref{eq:nEDM})  \cite{Crewther79}.
If the $\bar\theta$ term is present in QCD, then $\pi^0$ can obtain a VEV.  VEVs of $\pi^0$ and $\eta'$ break CP, and give   $|\overline{g_{\pi NN}}|\simeq\bar\theta/3$  \cite{KimRMP10}. The non-observation of nEDM put a limit on $|\bar\theta|$ as less than $10^{-10}$. For the class of calculable solutions, the so-called Nelson-Barr type weak CP violation is close to a solution \cite{Nelson,Barr}, but the limit $10^{-10}$ is difficult to realize.

\begin{figure}[t!]
\centerline{\includegraphics[width=0.4\textwidth]{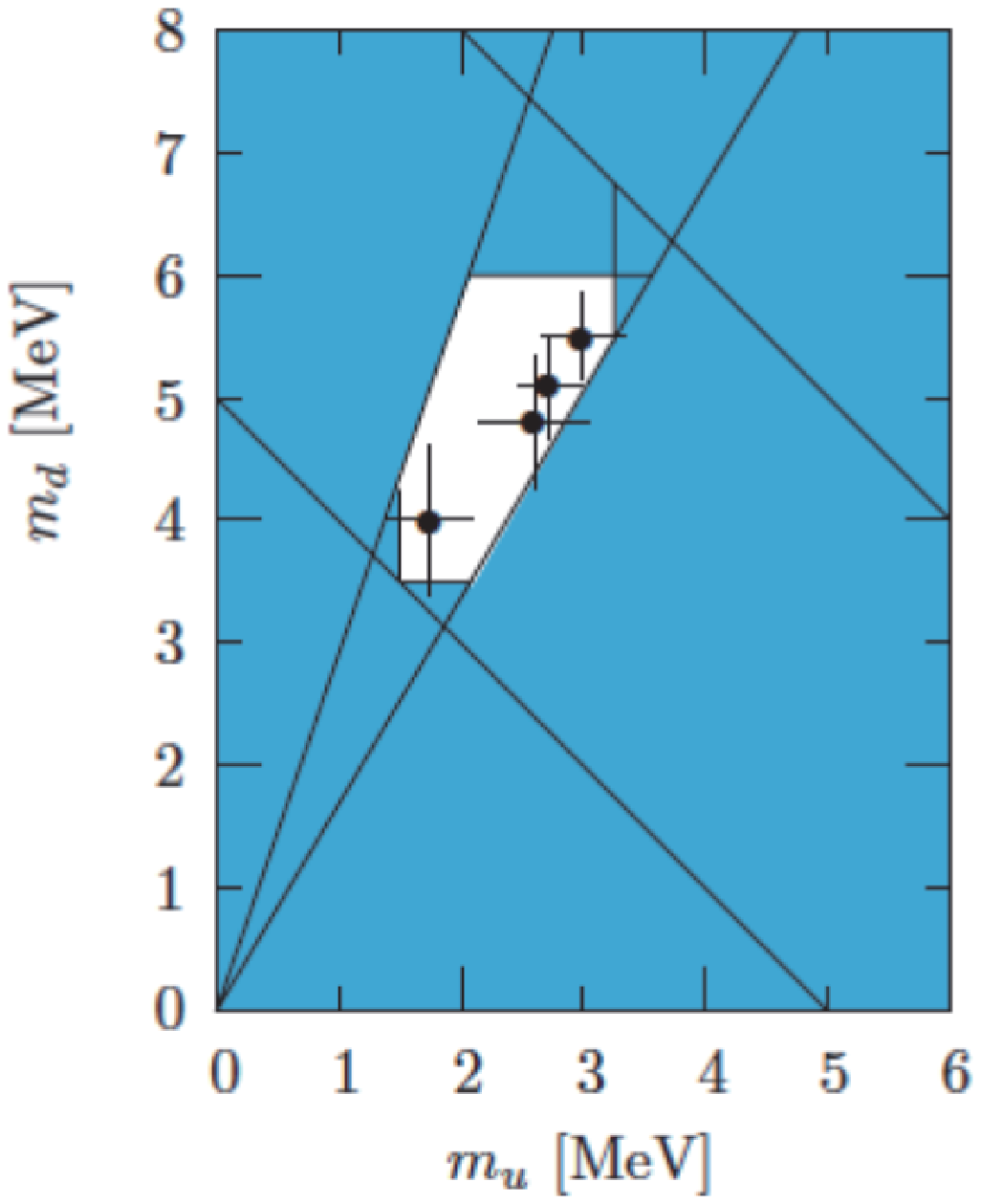}}
\caption{$m_d$ vs. $m_u$ \cite{Manohar14}.}\label{Fig:mu}
\end{figure}

For the massless up-quark solution, it seems not favored by the measured current quark masses as shown in Fig. \ref{Fig:mu} \cite{Manohar14}.
   
This leads us to a   brief historical introduction, eventually leading to the ``invisible'' axion \cite{KSVZ1,KSVZ2,DFSZ}. Pre- ``invisible'' axion developments are the following.

 At the time when the third quark family was not discovered, Weinberg
tried to introduce the weak CP violation in the Higgs potential. Satisfying the Glashow-Weinberg condition that up-type quarks couple to $H_u$ and down-type quarks couple to $H_d$ \cite{GW77}, he introduced many Higgs doublets.   Then, the weak CP violation introduced in the potential, with a discrete symmetry $\phi_I\to-\phi_I$,
\begin{eqnarray}
V_{\rm W}=\frac12\sum_I m_I^2\phi_I^\dagger\phi_I+\frac14 \sum_{IJ} \left\{a_{IJ} \phi_I^\dagger\phi_I\phi_J^\dagger\phi_J+b_{IJ}\phi_I^\dagger\phi_I\phi_J^\dagger\phi_J +(c_{IJ} \phi_I^\dagger\phi_I\phi_J^\dagger\phi_J+{\rm H.c.})
\right\}\label{eq:Weinberg}
\end{eqnarray}
Weinberg's necessary condition for the existence of CP violation is non-zero $c_{IJ}$ terms \cite{Weinberg76}. If one removes the $c_{IJ}$ terms, Peccei and Quinn (PQ) noticed that there emerges a global symmetry which is now called the  \UPQ~symmetry \cite{PQ77}. This is an example that keeping only a few terms among the discrete symmetry allowed terms in the potential produces a global symmetry.

With the PQ symmetry by removing the $c_{ij}$ term in Eq. (\ref{eq:Weinberg}) \cite{PQ77},  Weinberg and Wilczek at Ben Lee Memorial Conference noted the existence of a pseudoscalar, the PQWW axion 
\cite{PQWW77}, which was soon declared to be non-existent \cite{Peccei78}. This has led to calculable models discussed in
\cite{Beg78,Georgi78,Moha78,
Segre79,Langacker79}.

\section{The ``invisible'' axion}
However, a good symmetry principle of Fig. \ref{fig:Gross} is so an attractive framework, the PQ symmetry is re-introduced by a SM singlet field \cite{KSVZ1}, which has been named as ``invisible'' axion $a$. It was noted that the ``invisible'' axion was harmful in the evolution of the Universe
\cite{Preskill83}, which has been  later turned into a bonus after realizing that the CDM contribution was important in the evolution of the Universe \cite{Primack84}.

``Invisible'' can be made ``visible'' if one invents a clever cavity detector
\cite{Sikivie83}, which is used in many axion search labs now \cite{AxionLabs}.
\begin{figure}[t!]
\centerline{\includegraphics[width=0.85\textwidth]{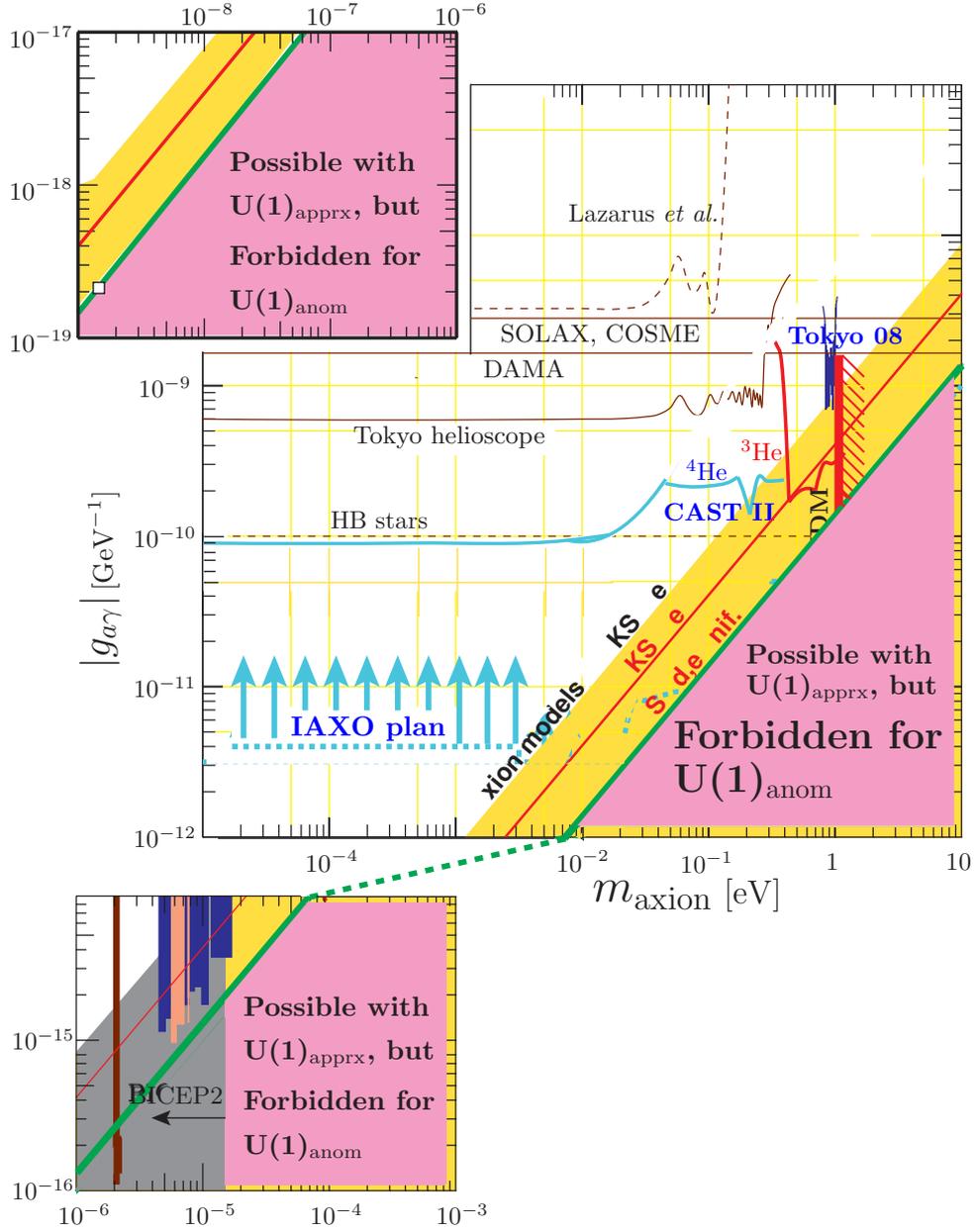}}
\caption{Axion detection bounds on the $c_{a\gamma\gamma}$ vs. $m_a$ plane with regions for several model parameters \cite{KimPLBagg15}. The white square on the upper left corner is the MI-axion point  \cite{KimNamPLB16}.}\label{Fig:AxData}
\end{figure}
There have been the cosmic experiment and  solar axion   search
experiments \cite{ADMX1}.  In Fig. \ref{Fig:AxData}, the current bounds on the ``invisible'' QCD axion search are shown.

An SU(2)xU(1) singlet housing the ``invisible'' axion gives the effective Lagrangian of $a$ as
 \begin{eqnarray}
 {\cal L}=c_1\frac{\partial_\mu a}{f_a} \sum_q\bar{q}\,\gamma^\mu\gamma_5\, q-\sum_q(\bar{q}_L\, m\, q_R\, e^{ic_2 a/f_a}  +\textrm{h.c.})+\frac{c_3}{32\pi^2 f_a}a\, G_{\mu\nu}\tilde{G}^{\mu\nu}
 \label{eq:invAxion}\\
 +\frac{c_{aWW}}{32\pi^2 f_a}a\, W_{\mu\nu}\tilde{W}^{\mu\nu}
 +\frac{c_{aYY}}{32\pi^2 f_a}a\, Y_{\mu\nu}\tilde{Y}^{\mu\nu} + {\cal L}_{\rm\, leptons},\nonumber
 \end{eqnarray}
where $\tilde{G}^{\mu\nu}, \tilde{W}^{\mu\nu}$, and $\tilde{Y}^{\mu\nu}$ are dual  field strengths of gluon, $W$, and hypercharge fields, respectively. 

It is a key question how the PQ symmetry is defined.  These couplings arise from the following renormalizable couplings,
  \begin{eqnarray}
  {\cal L}_{\rm KSVZ}& =&- f\overline{Q}_R Q_L+\textrm{h.c.} ,\label{eq:KSVZ}\\[0.5em]
  {V}_{\rm DFSZ}  &=&  -\mu_1^2 H_u^*H_u -\mu_2^2 H_d^*H_d+\lambda_1(   H_u^*H_u)^2+\lambda_2(   H_d^*H_d)^2+{\sigma\rm~terms} \label{eq:DFSZ}\\
  && (+M H_uH_d\sigma) +\lambda' H_uH_d\sigma^2+\textrm{h.c.}\nonumber
 \end{eqnarray}
 In the KSVZ model, the heavy quark $Q$ is introduced and the $f$-term Yukawa coupling is the definition of the PQ symmetry. In the DFSZ model, the $\lambda'$-term is the definition of the PQ symmetry.
Note, however, that  there must be a fine-tuning in the coefficient $\lambda'$ such that $\vew\ll f_a$ \cite{Dreiner14}. The axion-photon-photon couplings are listed in Table \ref{tab:Inv}.
\begin{table} 
\begin{tabular}{|r|c|} 
\hline &\\[-1em]
KSVZ:  $Q_{\rm em}$ ~&$c_{a\gamma\gamma}$\\\hline
$0$ ~~ &$-2$~ \\[0.3em]
$\pm \frac13$~~  &$-\frac{4}3$~  \\[0.3em]
$\pm \frac23$~~  &$ \frac{2}3$   \\[0.3em]
$\pm 1 $~~  &$4$  \\[0.3em]
$(m,m)$~&$-\frac{1}3$~  \\[0.2em] 
\hline
\end{tabular} 
\hskip 1cm
\begin{tabular}{|r c|c|} 
\hline &&\\[-1em]
DFSZ:  $(q^c$-$e_L)$ pair ~& Higgs &$c_{a\gamma\gamma}$\\\hline
\\[-1.35em]
&&  \\[-0.85em]
non-SUSY $(d^c,e)$ ~~ & $H_d$&$\frac23$~ \\[0.3em]
non-SUSY $(u^c,e)$ ~~  & $H_u^*$&$-\frac43$~ 
  \\[0.2em] 
GUTs  ~~ &  &$\frac23$~ \\[0.3em]
SUSY   ~~  &  &$\frac23$~ 
  \\[0.2em] 
\hline
\end{tabular} 
\caption{$c_{a\gamma\gamma}$ in the KSVZ and DFSZ models. For the $u$ and $d$ quark masses, $m_u=0.5\, m_d$  is assumed for simplicity. $(m,m)$ in the last row the KSVZ means $m$ quarks of $Q_{\rm em}=\frac23\,e$ and  $m$ quarks of $Q_{\rm em}=-\frac13\,e$.  SUSY in the DFSZ includes contributions of color partners of Higgsinos. If we do not include the color partners, \ie in the MSSM without heavy colored particles, $c_{a\gamma\gamma}\simeq 0$ \cite{BaeComm}.
}
\label{tab:Inv}
\end{table}

\begin{figure}[!t]
\centerline{\includegraphics[width=0.45\textwidth]{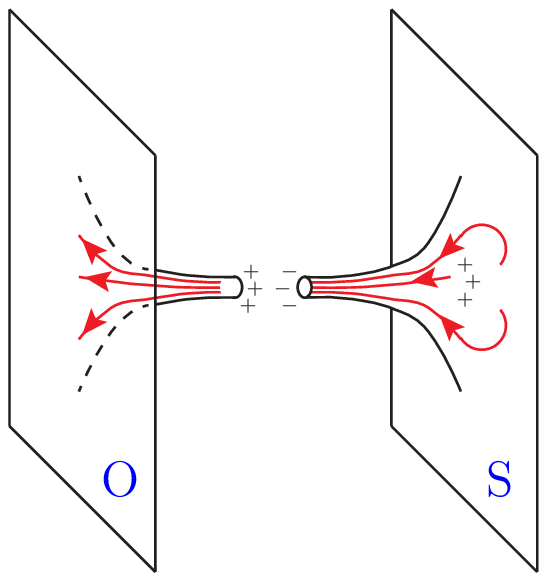}}
\caption{A wormhole connection to a shadow world.}\label{Fig:WormFlux}
\end{figure}

The fine-tuning problem in the DFSZ model is resolved in the supersymmetric(SUSY) extension of the model. There is no renormalizable term of the singlet superfield $\sigma$ with the SM fields. The leading term is the so-called Kim-Nilles term \cite{KN84},
 \begin{eqnarray}
W_{\rm KN}= \frac{1}{M}H_uH_d\sigma^2,
 \end{eqnarray}
 where $M$ is determined from a theory. It is shown in Table \ref{tab:Inv} as the SUSY $c_{a\gamma\gamma}$.
In Table \ref{tab:Inv}, $H_d$ and $H_u^*$ imply that they give mass to $e$.  GUTs and SUSY choose appropriate Higgs doublets and always give $c_{a\gamma\gamma}=\frac23$.

In the discussion of ``invisible'' axion, gravity effects was considered to be crucial. It started with the wormhole effects in the Euclidian quantum gravity \cite{Coleman88}. In fact, gravity equation with the antisymmetric tensor field $B_{\mu\nu}$ gives wormhole solutions  \cite{GS88}. This triggered the discrete symmetries allowable as subgroups of gauge groups  \cite{Krauss89}, and \UPQ~global symmetry needed for ``invisible'' axion
was considered to be problematic \cite{GravSpoil92}. It is euristically presented in Fig. \ref{Fig:WormFlux} for a flow of gauge charges from an observer O to a shadow world S through a wormhole. If he disconnects $S$, he recovers all gauge charges and concludes that no gauge charge is lost. This is because gauge charges carry hairs of flux lines. Thus, O confirms that gauge symmetries are not broken by wormholes. But, global symmetries do not carry flux lines and  quantum gravity does not allow global symmetries. Thus, ``invisible'' axion has the gravity spoil problem \cite{GravSpoil92}.
\begin{figure}[!b]
\hskip 1cm {\includegraphics[width=0.47\textwidth]{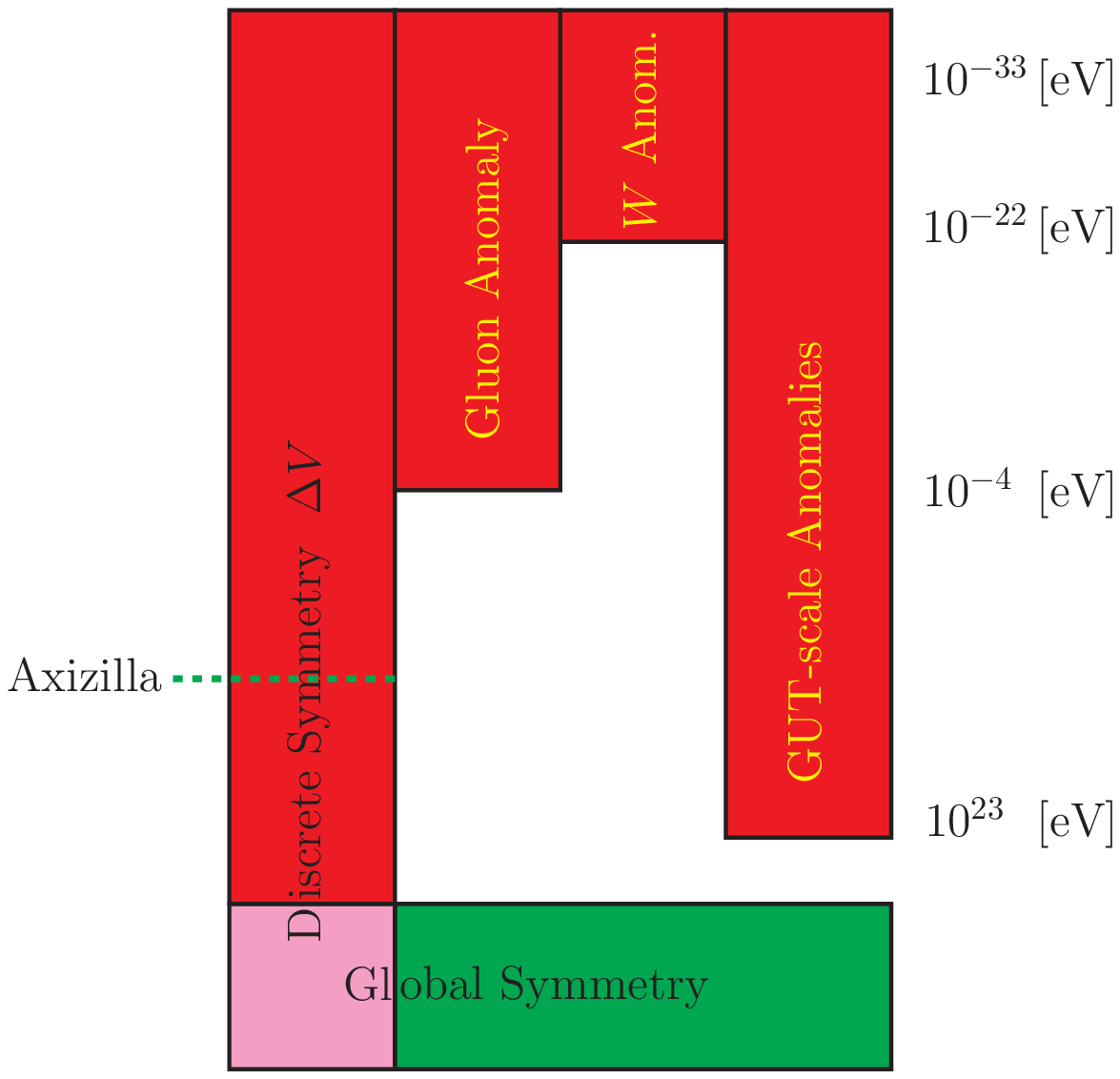}}\hskip 1cm {\includegraphics[width=0.33\textwidth]{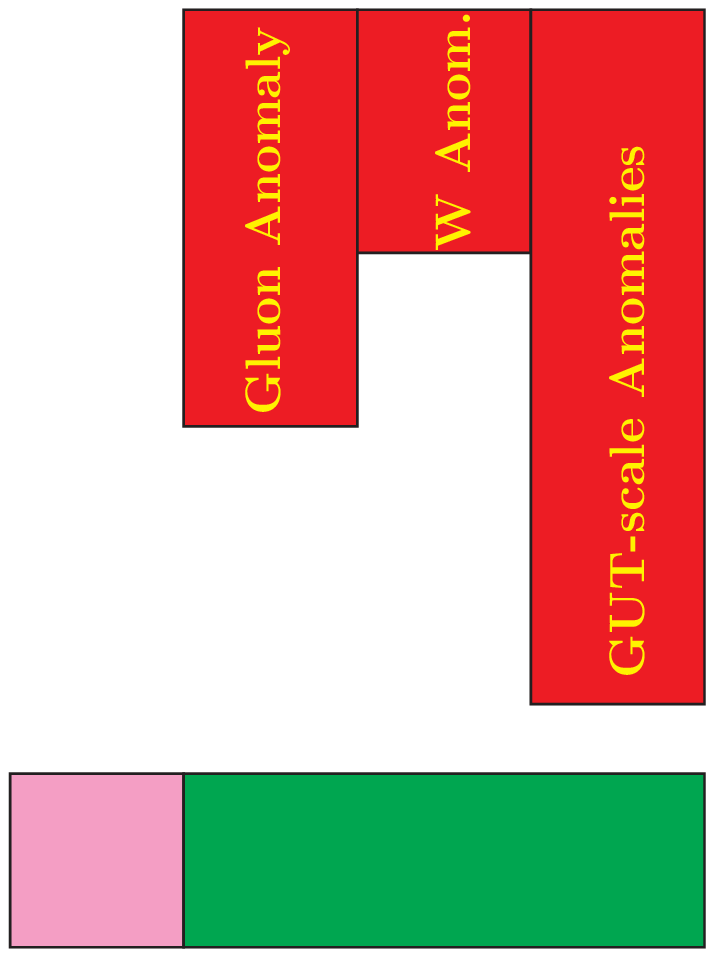}}\vskip 0.3cm
\centerline{\hskip 1cm(a)\hskip 6.7cm (b)}
\caption{Terms allowed in interactions. (a) Terms allowed by discrete symmetries (the most left column) and non-Abelian anomalies, and (b) terms allowed only by non-Abelian anomalies. Considering only terms in the lavender color, one finds   a global symmetry.}\label{fig:Terms}
\end{figure}

One may consider all terms allowed with some discrete symmetries, which is symbolized in the most left column of Fig. \ref{fig:Terms}\,(a). Discrete symmetries appear in most string compactification and the wormhole effects do not break these discrete symmetries. If only a few terms in all possible terms in $V$ are considered as symbolized in the lavender color in Fig. \ref{fig:Terms}, there may appear global symmetries. For example,  we can keep terms except the $c_{ij}$ terms  in Eqs. (\ref{eq:Weinberg}), which led to the \UPQ~symmetry as discussed before.
 These global symmetries are approximate and broken by terms in the red parts in Fig. \ref{fig:Terms}. The red part above the lavender symbolizes the terms in $V$.  
 One large violation of a global symmetry was suggested for heavy axions or axizillas at the TeV scale \cite{KimAxizilla}. The important one for ``invisible'' axion is the case of no $\Delta V$ as shown in  Fig. \ref{fig:Terms}\,(b). Then, the minimum of the potential is at $\bar{\theta}=0$ \cite{VW84}. If a term in $\Delta V$ is present, it must be sufficiently small such that the  ``invisible'' axion solution of the strong CP problem is intact.
 In any case, a PQ global symmetry can be obtained at least approximately from an ultra-violet completed theory. 
 
The well-known global symmetric operators are the effective neutrino mass term in the Lagrangian (instead of $V$) for U(1)$_L$ \cite{Weinberg79} and the Kim-Nilles term in SUSY for \UPQ,
  \begin{equation} 
{-\cal L}_{\nu\rm mass}=\frac{f_{ij}}{M}\ell_i\ell_j H_uH_u,~~~W_{\rm KN}=\frac{1}{M}\sigma_1\sigma_2 H_uH_d,\label{eq:def}
\end{equation}
where $\ell$'s are lepton doublets in the SM. Equations in (\ref{eq:def}) define the lepton number $L$ and the PQ quantum number $Q$,
\begin{eqnarray}
&&L(\ell\ell)=2,~\quad L(H_uH_u)=-2,\label{eq:Lnumber0}\\
&&Q(\sigma_1\sigma_2)=2,
~\quad Q(H_uH_d)=-2.\nonumber
\end{eqnarray}
For masses  of the SM  neutrinos, $\langle H_uH_u\rangle\ne 0$ is enough. The renormalizable interactions with right-handed neutrinos $N_R$ leading to Eq. (\ref{eq:Lnumber0}), $\ell_L H_u N_R$, are not needed for phenomenological neutrino masses at low energy.
   For the KN term also, renormalizable couplings, $\sigma_1 H_u X_{\rm doublet}$, $\sigma_2 H_d X'_{\rm doublet}, \overline{Q}_LQ_R\sigma_1,\cdots$, can give the term.

In this road toward detecting an ``invisible'' QCD axion, there has been a few theoretical development starting from an ultra-violet completed theory. The scale must be intermediate.  The model-indepent(MI) superstring axion \cite{MIaxion} is not suitable for this because the decay constant is about $10^{16\,}\gev$ \cite{ChoiKim85} which is the white square on the upper left corner in Fig. \ref{Fig:AxData}.
 
The question is, ``is it possible to obtain exact global symmetries?''
   
From string compactification, there is one way to make the ``invisible'' QCD axion to be located at the intermediate scale starting with an exact global symmetry,
\begin{equation} 
10^9\,\gev\le f_a\le 10^{11.5\,}\gev.
\end{equation}
It starts from the appearance of
an anomalous U(1) gauge symmetry in string compactification. In  compactifying the E$_8\times$E$_8'$ heterotic string, there appears an anomalous U(1)$_a$ gauge symmetry in many cases \cite{Anom86},
  \begin{equation} 
{\rm E}_8\times {\rm E}_8\to {\rm U(1)}_a\times \cdots
\end{equation}
Thus, the anomalous  U(1)$_a$ is belonging to a gauge symmetry of  E$_8\times$E$_8'$. In the original E$_8\times$E$_8'$ heterotic string, there is also the MI-axion degree $B_{\mu\nu}$. The gauge boson corresponding to this anomalous   \Uanom~ obtains mass by absorbing the MI-axion degree as its longitudinal degree. Therefore, the harmful MI-axion disappears, but not quite completely. Below the compactification scale of $10^{17\,}\gev$, there appears a global symmetry  which works as the PQ symmetry. This PQ symmetry can be broken by a SM singlet Higgs scalar(s), producing the ``invisible'' axion. So, this   ``invisible'' axion arises from an exact global symmetry  \Uanom, and is free from the gravity spoil problem because its origin is gauge symmetry.

Within this string compactification scheme, the axion-photon-photon coupling has been calculated   \cite{Kim88,KimPLBagg15,KimNamPLB16} and presented in Table \ref{tab:StAxion}.
   
\begin{table} 
\begin{tabular}{|l| r|c|} 
\hline  &&\\[-1em]
String:  &$c_{a\gamma\gamma}$  &Comments\\\hline\\[-1.35em]
&&  \\[-0.9em]
Ref. \cite{ChoiIWKim}  &  $-\frac13$~ &Approximate\\[0.3em] 
Ref. \cite{KimPLBagg15,KimNamPLB16} &  $\frac23$~ &Anom. U(1)\\[0.3em]
\hline
\end{tabular}
\caption{String model prediction of $c_{a\gamma\gamma}$.  In the last line, $c_{a\gamma\gamma}= (1-2\sin^2\theta_W)/\sin^2\theta_W$ with $m_u=0.5\, m_d$.
}
\label{tab:StAxion}
\end{table}

\section{CP and cosmology}
Consideration of CP in cosmological  
dark energy (DE), CDM,  and $\Delta B$ can be realated  CP symmetry and its violation. Quintessential axion \cite{QuintAxion}, QCD axion, and Sakharov's conditions are related to CP. $\Delta B$ needs
CP violation. In the Type-II leptogenesis, which will be given shortly, involves the weak CP violation. 
  
The axion solution of the strong CP problem is a cosmological solution.    QCD
axions oscillate with the CP violating
vacuum angle $\bar\theta$, but the average value is 0.  If the axion vacuum starts from $a/f_a=\theta_1\ne 0$, then the vacuum oscillates and this collective motion behaves like cold dark matter(CDM) as commented around Fig. \ref{Fig:AxData} \cite{BCM14}, for which a recent calculation energy density of coherent oscillation, $\rho_a$, is given in \cite{Bae08}.

\begin{figure}[!b]
\hskip 1cm {\includegraphics[width=0.67\textwidth]{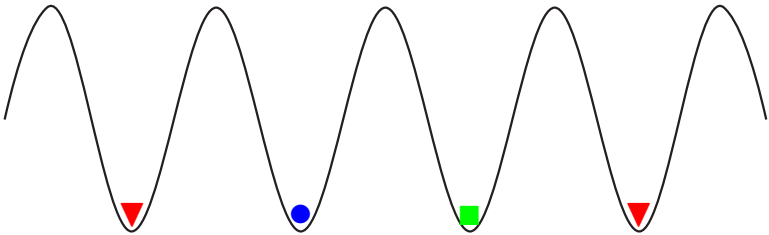}} 
 \caption{The axion vacuum with $\NDW=3$.}\label{fig:DWN3}
\end{figure}

The axion vacuum is identified by the shift of axion field by $2\pi\NDW f_a$,
\begin{equation}
a\to a+2\pi \,\NDW\,f_a.
\end{equation}
It is because the $\bar\theta$ term has the periodicity $2\pi$,
\begin{equation}
{\cal L}_{\bar\theta}=-\frac{a}{32\pi^2\,f_a}\int d^4x \,G^a_{\mu\nu}\tilde{G}^{a\,\mu\nu},~~a=a+2\pi f_a,
\end{equation}
while matter fields $\Phi$ may not have the periodicity of $2\pi f_a$, but only after $2\pi\NDW f_a$,
\begin{equation}
\Phi\to e^{i\bar\theta/\NDW}\Phi,~~
\bar\theta=\frac{a}{f_a}.
\end{equation}
The axion vacuum with $\NDW=3$ is shown in Fig. \ref{fig:DWN3}, where the red triangle vacuum is returning to itsef after going over three maxima. Between different vacua, there are domain walls.

\begin{figure}[!t]
\hskip 1cm {\includegraphics[width=0.47\textwidth]{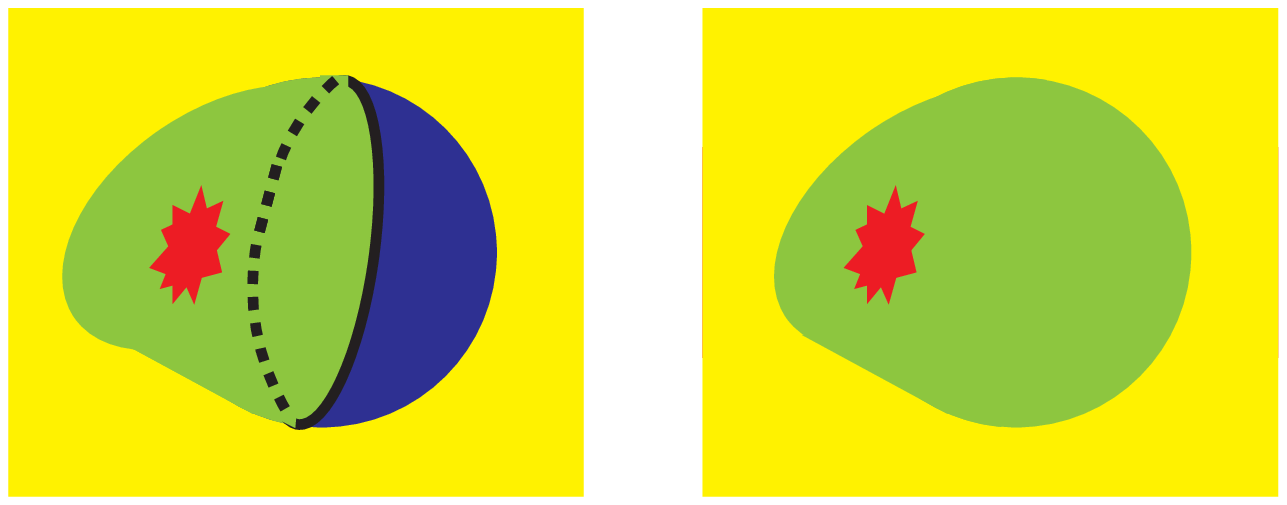}}\hskip 1cm {\includegraphics[width=0.33\textwidth]{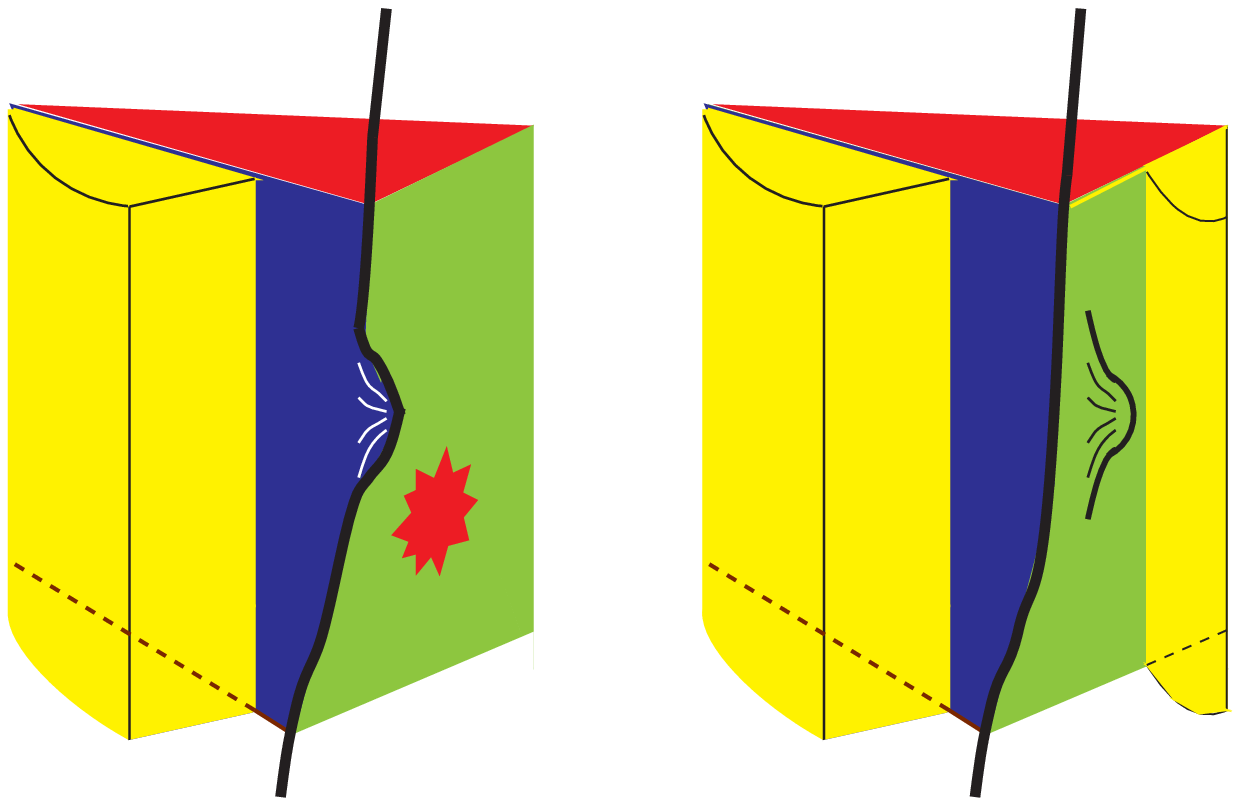}}\vskip 0.3cm
\centerline{\hskip 1cm(a)\hskip 6.7cm (b)}
\caption{Domain walls in $\Z_2$. (a) Domain wall ball, and (b) a cosmological scale domain wall.}\label{fig:DW2}
\end{figure}

\begin{figure}[!t]
\hskip 1cm {\includegraphics[width=0.67\textwidth]{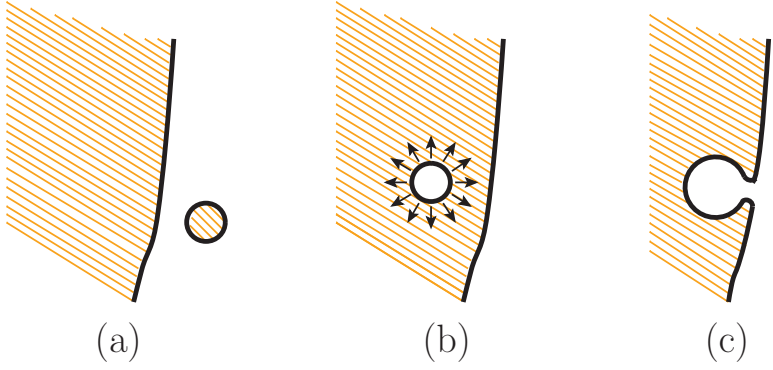}} 
 \caption{Domain walls in $\NDW=1$. (a) A DW ball is approaching to a horizon scale DW,  (b) the DW ball is punching the huge wall and expands with the light velocity, and (c) it eats up the huge wall \cite{Barr87}.}\label{fig:DW1}
\end{figure}

Topological defects of global \UG~produce an additional  axion energy density by the decay of string-wall system, $\rho_{st}$. Contribution of axionic string to energy density was known for a long time \cite{Davis86}. In addition, axionic domain walls carry huge energy density \cite{Vilenkin82, Sikivie82}. Because of the difficulty of removing comological scale domain walls for $\NDW\ge 2$ as sketched in Fig. \ref{fig:DW2}, it was suggested that
the axionic domain wall number should be 1 \cite{Sikivie82}.  If $\NDW=1$, the horizon scale walls can be annihilated as in Fig. \ref{fig:DW1} \cite{Vilenkin82,Barr87}. A small wall bounded by string collides with a horizon scale wall (a), and eats up the wall with the light velocity (b), eventually annihilating the horizon scale string-wall system (c).

Computer simulations use axion models with $\NDW=1$.
Three groups have calculated these which vary from O(1) to O(100),
\begin{eqnarray}
&&\textrm{Florida group:}~\textrm{O(1)} \cite{Florida01},\nonumber\\
&&\textrm{Cambridge group:}~\textrm{O(100)}   \cite{Cambridge94}, \label{eq:StContri}\\
&&\textrm{Tokyo group:}~\textrm{O(10)}    \cite{Kawasaki12}.\nonumber
\end{eqnarray}
A recent calculation for $\NDW=1$ models has been given  $\rho_{st}\sim  {\rm O(10)}\,\rho_a $ \cite{SekiguchiT}.

Therefore, it is important to realize axion models with $\NDW=1$. The KSVZ axion model with one heavy quark achieves $\NDW=1$. There are two other methods. One is identifying different vacua modulo the center number of the GUT gauge group   \cite{LS82}. Another important one is obtaining $\NDW=1$ by  the Goldstone boson direction \cite{Choi85,KimPLB16},
\begin{figure}[!t]
\hskip 1cm {\includegraphics[width=0.67\textwidth]{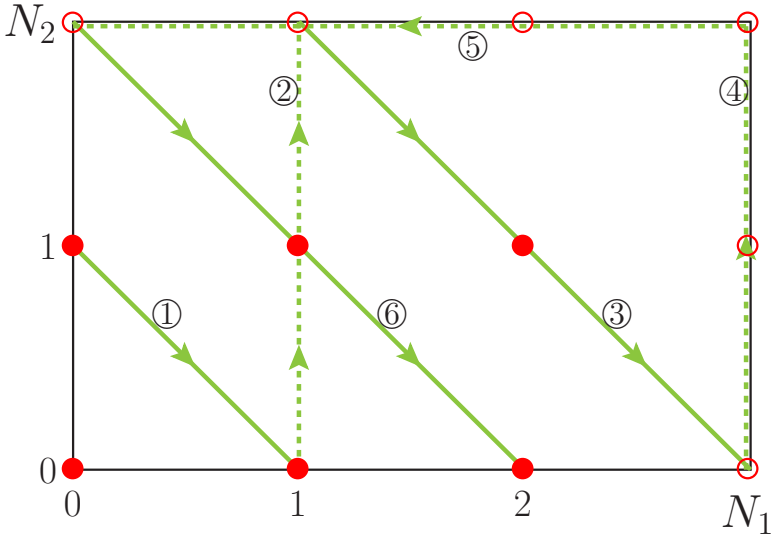}} 
 \caption{Vacua identification by Goldstone boson direction.}\label{fig:Goldstone}
\end{figure}
which is shown in Fig. \ref{fig:Goldstone}.
There are two degrees for the shifts, $N_1$ and $N_2$ directions. For the torus of $N_1=3$ and $N_2=2$ models, seemingly there are 6 vacua represended by red bullets in Fig. \ref{fig:Goldstone}. The Goldstone boson directions are shown as arrow lines and torus identifications are shown as dashed arrows. So, all six vacua are connected by one way or the other, and the  $N_1=3$ and $N_2=2$ model gives $\NDW=1$. One always obtain $\NDW=1$ if $N_1$ and $N_2$ are relatively prime. The reason that the ``invisible'' axion from \Uanom~ has $\NDW=1$ is because $N_{\rm from~E_8\times E_8'}=\rm large~integer$ but $N_{\rm MI~axion}=1$ \cite{Witten85,KimPLB16}, and  $N_{\rm from~E_8\times E_8'}$ and $N_{\rm MI~axion}$ are relatively prime.

There is another cosmological solution that the axionic string-wall system created at $f_a$ is allevated by another confining force at high temperature \cite{BarrKim14}.
The instantons of this force, acting below $f_a$, can generate an axion potential that erases the axion strings long before QCD effects become important, preventing QCD-generated axion walls from ever appearing.

Axionic string contribution is important if strings are created after PQ
symmetry breaking. On the other hand, with a high scale inflation this string contribution to energy density is important, as shown in Eq. (\ref{eq:StContri}). There can be a more important constraint if a large $r(=\,$tensor/scalar ratio) is observed. Two groups reported this constraint \cite{Gondolo14} after the BICEP2 report \cite{BICEP2}. Probably, this is the most significant impact of BICEP2 result on
$\NDW=1$ axion physics. The region is marked around $m_a\sim 71\,\mu$eV in Fig. \ref{Fig:AxData}.

\section{Type-II leptogenesis}

The mere 5\,\% of the energy pie,  mainly atoms composed of baryons of the Universe, belongs to the problem on the chiral representations of quarks in the GUT scheme \cite{Georgi79,KimICHEP1,KimJHEP15}.   In cosmology, it belongs to Sakharov's three conditions  \cite{Sakharov67},
\begin{itemize}
\item  $B$ number (or global quantum number) violation,
\item CP and C violation,
\item Out of thermal equilibrium.
\end{itemize}
For the last condition, we just make sure that the process proceeds
in non-equilibrium conditions. Usually, heavy particle decays proceed out of thermal equilibrium.  
The first one is the existence of baryon number  violating interaction implied in GUTs, and the second condition is on the CP and C violation.

Sphaleron processes at the electroweak scale changes baryon ($B$) and lepton($L$) numbers but do not change the combination $(B-L)$. If the sphaleron processes are 100\,\% effective in converting these global quantum numbers, a net baryon number is   $\Delta B|_{\rm after~sphaleron}\propto (B-L)$.
Thus, SU(5) GUT is problematic because  it conserves $B-L$ and cannot generate $(B-L)$ by the SU(5) processes. So, if one tries to use GUTs, choose one which violates $B-L$ such as in the SO(10) GUT.

But, a simple way is to work in the SM,
SU(3)$\times$SU(2)$\times$U(1), and introduce SM singlet fermions $N$ which are called neutrinos.  These $N$ particles are considered to be heavy.  This issue is the first condition on the global quantum number. If the SM singlets are present, there always appear neutrino masses, which implies that $L$ is broken. In defining the lepton number $L$, let us remember the Weinberg operator for neutrino masses \cite{Weinberg79,Minkowski77},
\begin{equation}
{\cal L}_{\nu\rm mass}=\frac{f_{ij}}{M}\ell_i\ell_j H_uH_u \label{eq:Wein}
\end{equation}
where  $\ell$ are left-handed lepton doublets, and gauge invariant indices are implied. $M$ is an effective mass in the non-renormalizable operator. The lepton number is defined with the left-handed SM doublets $\ell_L(\ni\nu_{e,\mu,\tau})$ carrying $L=+1$.
Non-zero neutrino masses break $L$. So, as far as $\langle H_u\rangle=0$, neutrinos do not obtain mass by the above operator. Therefore, $L$ can be properly defined below the scale of  $\langle H_u\rangle\ne 0$. One may argue that there is also $H_d$, which is exactly the reason that $H_u$ can carry a global quantum number, because both $H_u$ and $H_d$ can together define a gauge charge $Y$ and a global charge $L$.  Two $\ell$'s carry $+2$. However, who cares about renormalizable terms very importantly at low energy?
In cosmology, however, it is important.  
In cosmology, lepton number of $H_u$ is defined by
\begin{itemize}  
\item[]{\bf Type-I}:~~~ $L(H_uH_u)=0$,
\item[]{\bf Type-II}:~~ $L(H_uH_u)=-2$. 
\end{itemize}  
The quantum numbers of Type-I   \cite{FY86} and Type-II \cite{CoviKim16} leptogeneses are suggested by the following renormalizable term, and the quantum numbers of $L$,
\begin{eqnarray}
  \ell_L~~~H_u~~~~~N_L&&\\
{\rm Type-I}:~  +1,~~~0,~~~-1&&\\
{\rm Type-II}:~  +1,~ -1,~~~~~0&& 
\end{eqnarray}
The early idea was to define a right(left)-handed $N$ with lepton number $L=+1(-1)$ \cite{FY86}. In fact, the definition of lepton number, related to neutrino masses, is a combination of defining the lepton number  of the up-type Higgs doublets together with that of $N$. 

\begin{figure}[t!]
  \begin{center}
   \includegraphics[width=0.55\textwidth]{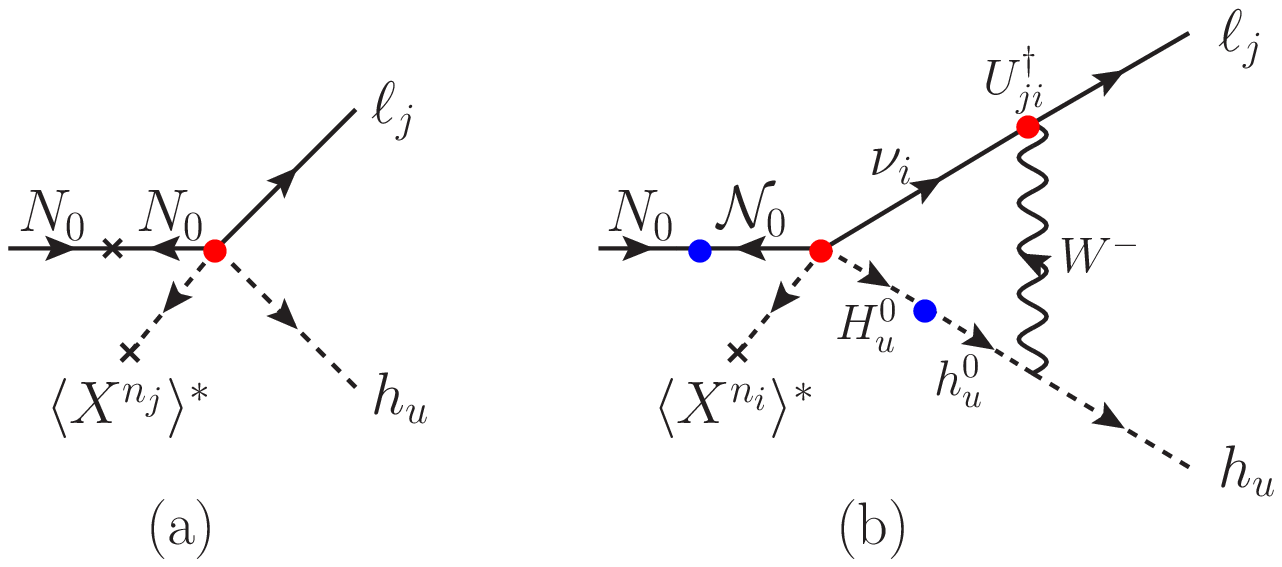}\\ ~\vskip 0.3cm
  \includegraphics[width=0.65\textwidth]{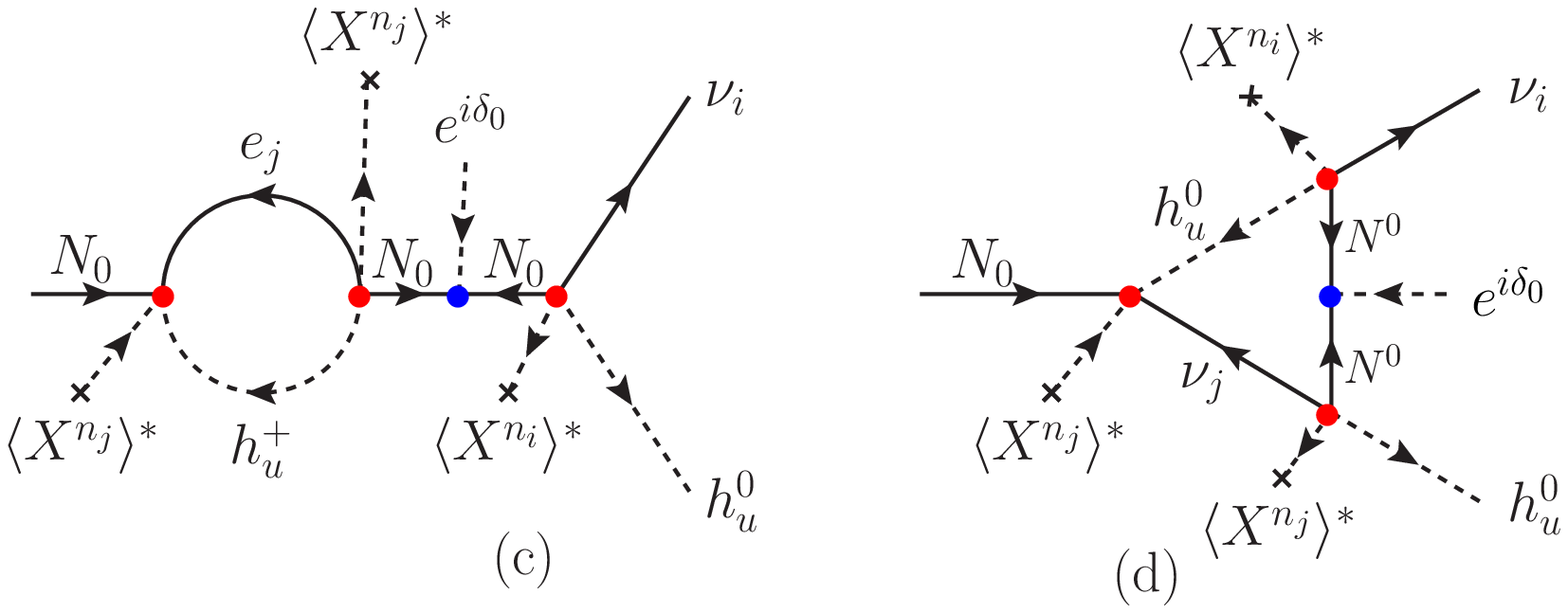}\hskip 1.2cm \includegraphics[width=0.2\textwidth]{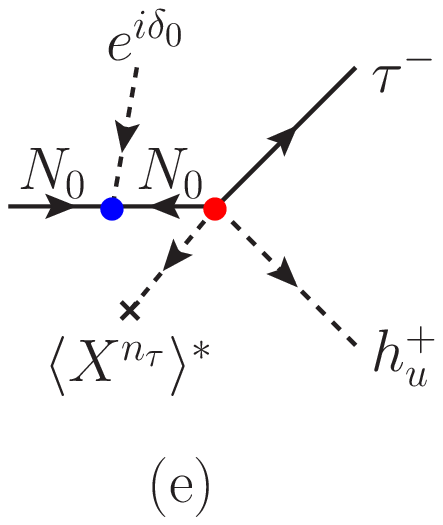}
  \end{center}
 \caption{The Feynman diagrams interfering in the $N$ decay: (a) the lowest order diagram in Type-II, (b) the $W$ exchange diagram, (c) the wave function renormalization diagram, and (d)  the heavy neutral lepton exchange diagram.  There exist similar ${\cal N}$-decay diagrams.   In all figures, the final leptons can be both charged leptons and neutrinos. 
 The lepton number violations are inserted with blue bullets and phases are inserted at red bullets. (a) and (b) interfere.  (c) and (d) give a vanishing contribution in $N_0$ domination with one complex VEV.} \label{fig:IntCP}
\end{figure}

`Type-II leptogenesis' is proposed  under a different definition on $L$ from that of Type-I and the electroweak symmetry breaking at high temperature. There exists a finite region of parameter space in the multi-Higgs model that the electroweak symmetry is broken at high temperature  \cite{Senj79}. In SUSY, the electroweak symmetry breaking at high temperature is more probable since the temperature dependent terms in $V$ are SUSY breaking terms.

In Fig. \ref{fig:IntCP}, we present diagrams appearing in Type-I and Type-II leptogeneses. The SM Higgs doublet is $h$, and an inert Higgs doublet is $H$. (e), (c), and (d) give Type-I where both the $L$ violation and CP violation appear in blue bullets. In Type-I, CP violation is introduced at the mass term of $N$, needing  at least two $N$'s for the interference of  Figs. \ref{tab:DefQno}\,(e), (c), and (d). So, if there is a hierarchy of masses such as $m_{N_0}/m_{N_1}\ll 1$, then the lepton asymmetry is suppressed by that factor.

\begin{table} 
\centerline{\begin{tabular}{|l| ccc|} 
\hline  &&&\\[-1em]
  &$\ell_L$ &$h_u\,(H_u)$ & $N$\\ \hline\\[-1.4em]
&&&  \\[-1em]
Type-I leptogenesis  &  ~$+1$~ &~$0$~&~$-1$~\\[0.3em] 
Type-II leptogenesis &  ~$+1$~ &~$-1\,(-2)$~ &$0$~\\[0.3em]
\hline
\end{tabular}}
\caption{Definition of lepton numbers, in Type-I and Type-II leptogeneses.}
\label{tab:DefQno}
\end{table}

\begin{table} 
\centerline{\begin{tabular}{|c| c ccccc|} 
\hline  &&&&&&\\[-1em]
  &~$\ell_L$~ &$H_u$&$H_d$ & $h_{u}$&$N$ & ${\cal N}$\\ \hline\\[-1.4em]
&&&&&&  \\[-1em]
 Type-II leptogen. &  ~$+1$~ &~$-2$~&~$+2$~ &~$-1$~&~$0$~&~$+1$~\\[0.3em]
 VEV &  ~$\times$~ &~\textrm{inert}~ & $ \textrm{inert}$  & ~$v_{\rm ew}\,s_\beta$ &~$\times$~&~$\times$~\\[0.3em]
\hline
\end{tabular}}
\caption{Definition of lepton numbers  in Type-II leptogeneses. We introduced an inert Higgs $H_u$ carrying $L=-2$ with zero VEV and singlet leptons ${\cal N}$ carrying $L=+1$.}
\label{tab:Type2}
\end{table}
In Fig.  \ref{fig:IntCP}, (a) and (b) give Type-II, and   CP violation is introduced by the PMNS matrix. Thus, we must have $W$ boson is not massless, \ie SU(2)$\times$U(1) symmetry is broken at the high temperature. Indeed, such possibility has been suggested long time ago \cite{Senj79}.   One can define the {\it lepton number} $L$ as shown in Table \ref{tab:DefQno}. In Type-II, we detail the lepton numbers in Table \ref{tab:Type2}. Here, we need an inert Higgs doublet $H_u$ carrying $L=-2$  and singlet leptons ${\cal N}$ carrying $L=+1$.  Here, the lepton number violation  appears in blue bullets and the CP violation appears in red bullets. Anyway, the fields $H_{u,d}$ and ${\cal N}$ introduced at high energy scale are not visible at low energy scale. We  introduce interactions
\begin{eqnarray}
&N_0\,\ell_L h_u,~{\cal N}_0\,\ell_L H_u,~N_0\,N_0,~H_uH_d,\cdots\\
&h_u^*H_u, ~N_0\,{\cal N}_0,~{\cal N}_0{\cal N}_0, \cdots 
\end{eqnarray}
where the first line conserves $L$ and  the second line violates $L$.

Within this framework, we calculated the lepton asymmetry which turned out to be
consistent with the fact that in these diagrams 
the lepton violation is on the left side of the cut diagram as discussed in \cite{Adhikari:2001yr}. 
The lepton asymmetry we obtain is a form,
\begin{eqnarray}
\epsilon_L^{N_0}(W)&
\approx \frac{\alpha_{\rm em}}{2\sqrt{2}\sin^2\theta_W}  \frac{\Delta m_h^2 }{m_0^2}\\
&\cdot\sum_{i,j}{\cal A}_{ij}\sin[(\pm n_P+n'-n_i+n_j)\delta_X],
\label{eq:asymmExp4}
\end{eqnarray}
where ${\cal A}_{ij}$ are given by Yukawa couplings, $n_P,n',n_i,n_j$ are integers, and we assumed that only one phase $\delta_X$ appears in the full theory.  Two independent $n'$ are Majorana phases multiplied to the PMNS matrix.
Using the sphaleron calculation of  \cite{D'Onofrio:2014kta}, we obtain
\begin{eqnarray}
\frac{\Gamma_{\rm sph}^{\rm broken}}{T^3 H(T)} = 
\kappa \alpha_W^4 \left(\frac{ 4\pi k}{g_W} \right)^7 e^{- 1.52 k \frac{4\pi}{g_W}} \sqrt{\frac{90}{\pi^2 g_*}} 
\frac{M_P}{T} \geq 1,
 \end{eqnarray}
leading to the constraint
\begin{eqnarray}
C_2 C_3 \sin\delta_c + C_1 S_2 S_3 \sin (\delta_c +
\dell) \simeq  2.4 \times 10^{-2},
 \end{eqnarray}
where $\delta_c$ is a Majorana neutrino phase. So, there is an enough parameter space to allow an acceptable $\Delta L$.
\section{Conclusion}

My talk is centered around  CP symmetry/violation and its cosmological effects.  A few emphases were: 
 (1) the Jarlskog determinant $J$ is $|{\rm Im\,} V_{31} V_{22} V_{13}|$ in the KS form,  
 (2) it is shown that the Jarlskog determinant $J$ is almost maximum with the current determination of quark (real) mixing angles, (3)  why we need ``invisible'' axions for a solution of strong CP problem, (4)  we commented some cosmological problems, (5) the ``invisible'' axion from an exact \UG~with $\NDW=1$ is possible  if it arises from \Uanom~ in string compactification, and (6) we discussed also the recent idea of Type-II leptogenesis.   

\paragraph{Note added.} After the talk, there appeared some relevant papers  \cite{UanomFl,Dvali16,DiscBUp}.


\end{document}